\documentclass[aps,twocolumn,showpacs,superscriptaddress]{revtex4}
\usepackage{amsmath}
\usepackage{amssymb}
\usepackage{epsfig}
\usepackage{color}
\usepackage[colorlinks]{hyperref}
\hypersetup{colorlinks,citecolor=red,linkcolor=blue,urlcolor=blue}
\usepackage{graphicx,amsmath}
\UseRawInputEncoding \input
\begin{document}

\title{Scaling behavior of superconductors}
\author{V. R. Shaginyan}\email{vrshag@thd.pnpi.spb.ru}
\affiliation{Petersburg Nuclear Physics Institute, NRC Kurchatov
Institute, Gatchina, 188300, Russia}\affiliation{Department of
Physics, Clark Atlanta University, Atlanta, GA 30314,
USA}\author{A. Z. Msezane} \affiliation{Department of Physics,
Clark Atlanta University, Atlanta, GA 30314, USA}
\author{S. A. Artamonov} \affiliation{Petersburg Nuclear Physics Institute,
NRC Kurchatov Institute, Gatchina, 188300, Russia}

\begin{abstract}
In our brief review, we will consider the general universal scaling
properties of superconductors. The physics of superconductors,
represented by both conventional and unconventional
superconductors, has been the main topic of high-$T_c$
superconductor physics for over thirty years, revealing some of the
properties of high-$T_c$ (or unconventional) superconductors.
Scaling relationships lead to the identification of fundamental
laws of nature and reveal the essence of superconductor physics.
Advances in experimental technology allow us to collect important
data, which in turn allow us to make definitive statements about
the physical processes underlying strongly correlated Fermi
systems. We analyze the scaling of the condensation energy
$E_{\Delta}$ divided by $\gamma$, $E_{\Delta}/\gamma\simeq
N(0)\Delta_1^2/\gamma$, where $N(0)$ is the density of states,
$\Delta_1$ is the maximum value of the superconducting gap and
$\gamma$ is the Sommerfeld coefficient. We show that the universal
$E_{\Delta}/\gamma\propto T_c^2$ scaling is equally applicable to
both conventional and unconventional high-$T_c$ superconductors. We
also analyze interesting experimental data demonstrating unique
behavior of unconventional superconductors, which radically
distinguishes them from conventional superconductors. To this end,
we consider the differential resistance (or conductivity) collected
under the application of magnetic field on the archetypical HF
metal $\rm CeCoIn_5$ that represents unconventional superconductor.
We show that the observed scattering rate is explained by the
occurrence of flat bands, while the so-called Planck limit arises
by chance, since normal metals exhibit the same scattering rate
behavior. We analyze experimental facts that reveal the same
universal properties of both unusual and ordinary superconductors
and theoretically explain  that $d\rho(T)/dT\propto\lambda^2_D$ and
$\rho_{s0}\propto T_c\sigma(T_c)$. Where $\rho$ is the resistivity,
$T$ is the temperature, $\lambda_D$ is the zero-$T$ penetration
depth, and $\rho_{s0}$ is the superconducting electron density. We
analyze the Homes' law and provide a theoretical explanation of its
general scaling applicable to superconductors. Overall, these
scaling relationships lead to the identification of fundamental
laws of nature and reveal the essence of superconductor physics.
All these observations support the theory of fermion condensation.
Our theoretical results agree well with a body of diverse and
seemingly unrelated experimental facts. They show that the
topological fermion condensation quantum phase transition
generating flat bands is an intrinsic property of strongly
correlated Fermi systems and can be considered as a universal agent
explaining their basic physics.
\end{abstract}


\maketitle

\section{Introduction}\label{I}

The common belief is that strongly correlated, or unconventional,
superconductors have nothing in common with conventional
superconductors. Indeed, unconventional superconductors are metals
with flat bands \cite{catal} in the absence of quasiparticles, see
e.g. \cite{scn}. On the other hand, experimental facts show that
both types of superconductors have common properties, have
quasiparticles and exhibit common scaling behavior of the scaled
condensation energy $E_{\Delta}/\gamma$, see, for example,
\cite{prb2015,prlq,mat,npbq,Narit}. The physics of unconventional
superconductors represented by strongly correlated metals, being
the mainstream topic of uncommon superconductors for more then
thirty years, manifests some features of common superconductors.
Recent advancements in experimental techniques permit to collect
important data, which, in turn, allow one to make the conclusive
statements about the underlying physics of strongly correlated
Fermi systems. These systems exhibit the non-Fermi-liquid (NFL)
properties. One of the physical sources of NFL behavior is unusual
property, which may occur in such systems. Namely, at $T=0$, a
portion of their excitation spectrum becomes dispersionless, giving
rise to flat bands. The presence of a flat band changes the
topology of the Fermi surface, creating a new class of Fermi liquid
\cite{vol}. This indicates that the system is close to a special
quantum critical point, which separates normal Fermi liquid and
that with the fermion condensate (FC). This quantum critical point
is coined as the topological fermion-condensation quantum phase
transition (FCQPT).

An essential aspect of behavior of a system hosting a flat band is
that the application of magnetic field restores its normal
Fermi-liquid properties. Taking into account the topological FCQPT
forming flat bands, we analyze the scaling of the condensation
energy $E_{\Delta}$ divided by $\gamma$, $E_{\Delta}/\gamma\simeq
N(0)\Delta_1^2/\gamma$, where $N(0)$ is the density of states,
$\Delta_1$ is the maximum value of the superconducting gap and
$\gamma$ is the Sommerfeld coefficient. We show that the universal
scaling of $E_{\Delta}/\gamma\propto T_c^2$ applies equally to both
conventional superconductors and unconventional superconductors
with high $T_c$ \cite{prb2015}. We study experimental facts that
reveal the same universal properties of both unconventional and
conventional superconductors and theoretically explain that
$d\rho(T)/dT\propto\lambda^2_D$ and $\rho_{s0}\propto
T_c\sigma(T_c)$  \cite{mac,epl22,lin,donald}. Where $\rho$ is the
resistivity, $T$ is the temperature, $\lambda_D$ is the penetration
depth at $T=0$, and $\rho_{s0}$ is the superconducting electron
density. Our consideration is based on both facts: Bogoliubov
quasiparticles act in conventional and unconventional
superconductors, and the corresponding flat band is deformed by the
unconventional superconducting state. Our theoretical analysis is
also based on both the experimental paper that examines a
representative subset of cuprates under optimal doping without any
pseudogap \cite{prb2015} and our theoretical papers
\cite{ms,phys_rep,cryst24}. Our discussion also applies to
graphene, since it has flat bands that shape its typical behavior
observed in other unconventional superconductors, see, for example,
\cite{vol15,catal,scr,khod20,epl22,book_20,nphys19}. As a result,
our theoretical observations based on the fermion condensation
theory are in good agreement with experimental facts. Note that the
common opinion suggests that flat bands are not deformed by the
corresponding superconducting state, see, for example,
\cite{Aror,Tian}, which contradicts experimental facts, as we will
see below.

Within the FCQPT formalism, our review paper considers recent
exciting experimental observations of universal scattering rate
related to linear temperature dependence of resistivity in a large
number of strongly correlated Fermi systems as well as normal
metals. We show that the observed scattering rate is explained by
the emergence of flat bands formed by the topological FCQPT, while
the so called Planckian limit occurs accidentally since the normal
metals exhibit the same scattering rate behavior. As an example, we
also analyze recent challenging experimental data on differential
resistivity (or conductivity) collected under the application of
magnetic field on the archetypical HF metal $\rm CeCoIn_5$ that
represents unconventional superconductor. We also analyze
experimental facts that reveal coinciding universal properties of
both uncommon superconductors and common ones and explain these
facts. We show that these observations support the FC theory. Our
theoretical results are in good agreement with corps of different
and seemingly unrelated experimental facts. They show that the
FCQPT is an intrinsic property of strongly correlated Fermi systems
and can be viewed as the universal agent explaining their core
physics.

The flat band problem could have been solved many years ago within
the Landau's Fermi liquid (LFL) theory \cite{lanl}. The theory is
based on the assumption that the energy functionals $E_0[n({\bf
p})]$ in the functional space $[n]$ of quasiparticle distributions
$n({\bf p})$ at zero temperature $T=0$ are in the functions $[n]$
taking the values 0 and 1. This theory is based on the assumption
that the single particle spectrum of quasiparticles of a normal
Fermi liquid is similar to the spectrum of an ideal Fermi gas,
differing from the latter in the value of the effective mass $M^*$.
At temperature $T=0$ in a homogeneous isotropic substance, the
distribution of quasiparticles of the ground state LFL is a Fermi
step function $n_F(p)=\theta (p-p_F)$. Quasiparticles fill the
Fermi sphere up to the same Fermi momentum $p_F$, that is
$p_F=(3\pi^2\rho)^{1/3}$,  where $\rho$ is the number density
\cite{lanl}. This assumption remains valid as long as the necessary
stability condition is satisfied:
\begin{equation}\label{nsc}
\delta E_0=\int (\varepsilon[{\bf p},n( {\bf p},T=0)] -\mu)\delta
n({\bf p},T=0)\frac{d^3p}{(2\pi)^3}>0.
\end{equation}
Here $\varepsilon[{\bf p},n({\bf p})]={\delta E_0[n]/\delta n({\bf
p})}$ is the quasiparticle energy, $n({\bf p})$ is the
quasiparticle distribution function, and $\mu$ is the chemical
potential. The condition \eqref{nsc} requires that the change in
the Landau functional $E_0[n]$ \cite{lanl}  for any admissible
variations of $n_F(p)$ is preserved. Thus, it is the violation of
the condition specified by Eq. \eqref{nsc} that leads to a
restructuring of the distribution of $n_F({\bf p})$. The
distribution function of quasiparticles $n({\bf p})$ is limited by
the Pauli exclusion principle $1\geq n({\bf p})\geq 0$. Thus, we
find two classes of solutions of Eq. \eqref{nsc}. The first general
class is defined as $\delta n({\bf p})=0$ with $n({\bf p})=0$ or
$n({\bf p})=1$, that is $n({\bf p})=n_F({\bf p})$ \cite{lanl}. The
second class, representing flat bands, is
\begin{equation}\label{7**}
\varepsilon(p)=\mu;\,\, {\rm if}\,\, 1>n_0({\bf p})>0\,\, {\rm
in}\,\, p_i<p<p_f.
\end{equation}
Which takes place if $n_0({\bf p})$ becomes $1>n_0({\bf p})>0$ in
region $p_i<p_F<p_f$ \cite{ks,ksk,phys_rep,vol,Volovik}. The
presence of flat bands signals that the system is close to a
special quantum critical point (QCP), separating normal Fermi
liquid and that with fermion condensate (FC)
\cite{ks,vol,khod95,phys_rep,book,book_20,meln17}. This QCP is
coined as the topological fermion-condensation quantum phase
transition (FCQPT), leading to FC formation \cite{phys_rep}. Flat
bands are observed in many strongly correlated Fermi systems
\cite{catal} that first arose as a mathematical exercise
\cite{ks,ksk} and are now a rapidly expanding and dynamic field
with countless applications, see e.g.
\cite{catal,book_20,phys_rep,Volovik,prl20,bern,bern_prl}. For
example, in case of high-$T_c$ superconductors the critical
temperature \cite{ks,ksk,Volovik,phys_rep,prl20,bern,bern_prl}
\begin{equation}\label{TC}
T_c\propto \Delta_1\propto \lambda_0,
\end{equation}
rather than being $T_c\propto\exp{(-1/gN(0))}$, where $\lambda_0$
is the superconducting coupling constant, $\Delta_1$ is the maximum
value of the superconducting gap and $N(0)$ is the density of
states at the Fermi surface \cite{bcs,til,genn}. Thus, it can be
assumed that high-temperature superconductors, which belong to
strongly correlated Fermi systems and exhibit non-Fermi-liquid (NF)
behavior, cannot have anything in common with conventional
superconductors. On the other hand, the condensation energy of both
unconventional and conventional superconductors exhibits a
universal scaling behavior: $E_{\Delta}/\gamma\simeq
N(0)\Delta_1^2/\gamma\propto T_c^2$, as it follows from
experimental facts \cite{prb2015}.

In our review we analyze both unconventional high-$T_c$
superconductors and conventional ones, and demonstrate that both of
them exhibit the common universal scaling of the condensation
energy $E_{\Delta}/\gamma$, $E_{\Delta}/\gamma\simeq
N(0)\Delta_1^2/\gamma$. We demonstrate  that the universal scaling
of $E_{\Delta}/\gamma\propto T_c^2$ applies equally to conventional
and unconventional high-$T_c$ superconductors. Our results are in
good agreement with experimental facts \cite{prb2015}. This
observation suggests that the FC superconducting state is Bardeen-
Copper-Schrieffer-like (BCS) and suggests the fundamental
applicability of the BCS formalism to describe some properties of
the superconducting state, as predicted in \cite{phys_rep,jetplbq}.
This observation allows us to give theoretical explanation of
Homes' law \cite{homes},
\begin{equation}\label{home}
\rho_{s0}\propto T_c\sigma(T_c).
\end{equation}
This general property applies to both conventional and
unconventional superconductors, allowing us to consider both types
of superconductors on an equal footing. Where $\sigma$ is the
conductivity, $T_c$ is the critical temperature of the
superconducting phase transition, and $\rho_{s0}$ is the density of
superconducting electrons at $T=0$. We also consider observations
of low-temperature linear resistivity $\rho(T)\propto T$, which
relates the slope of linear-$T$-dependent resistivity to the London
penetration depth $\lambda_D$ indicating a universal scaling
property
\begin{equation}\label{hzo}
\frac{d\rho}{dT} \propto \lambda_D^2
\end{equation}
for a large number of strongly correlated hight-temperature
superconductors \cite{lin}. This scaling relation spans several
orders of magnitude in $\lambda_D$, attesting to the robustness of
the empirical law \eqref{hzo}, where $\lambda_D$ is the zero-$T$
penetration depth.

We remark that the quasi-classical physics is still applicable to
describe the $T$-linear dependence of resistivity of strongly
correlated metals since flat bands, forming the quantum
criticality, generate transverse zero-sound mode with the Debye
temperature $T_D\sim 10 $ mK \cite{quasi,khod2012}. At $T\geq T_D$
the mechanism of the $T$-linear dependence is the same both in
ordinary metals and strongly correlated ones, and is represented by
electron-phonon scattering. Therefore, it is the electron-phonon
scattering at $T\geq T_D$ leads to the near material-independence
of the lifetime $\tau$, which is involved in the relation as
$1/(\tau T)\sim k_B/\hbar$. As a result, we describe and explain
recent exciting experimental observations of universal scattering
rate related to linear-temperature resistivity in a large number of
both strongly correlated Fermi systems and ordinary metals
\cite{bruin,legr,cao,nakaj}. We show that the observed scattering
rate is explained by the emergence of flat bands formed by the
topological FQCPT rather than by the so-called Planckian mechanism.
For the Planckian limit for the scattering rate  can hardy occur
for ordinary metals, see Fig. \ref{Sc1}. Apart from above, we
analyze specific features of strongly correlated metals like
differential resistivity (or conductivity) collected under the
application of magnetic field on the archetypical HF metal $\rm
CeCoIn_5$ \cite{steg11,steg17,graph}. We also consider outstanding
experimental observation of the Leggett theorem \cite{leg}
violation in overdoped high-$T_c$  copper oxides \cite{bosovic}.
Namely, it turns out that in these compounds at $T \to 0$ the
density of superconducting electrons $n_s$ can become much less
than the total electronic density$n_{el}$, $n_s<<n_{el}$.

In Section \ref{II} we consider the superconducting systems with
the  fermion condensation state.

In Section \ref{III} we analyze common scaling of the condensation
energy $E_{\Delta}/\gamma$ of both conventional and high-$T_c$
superconductors.

The following Sections \ref{IV}, \ref{cop} and \ref{asc} are
devoted to the specific properties of unconventional
superconductors that distinguish them from conventional ones.

Section \ref{IV} is devoted to linear temperature dependence of
resistivity and universal scattering rate and to consideration of
the Planckian limits.

In Section \ref{cop} the superfluid density of overdoped copper
oxides is analyzed.

In Section \ref{asc} the asymmetrical conductivity of HF metals is
considered.

Section \ref{unpr} is devoted to the universal properties of
heavy-fermion metals, conventional and non-conventional
superconductors.

In Section \ref{conc} the main results of our review are collected.

\section{Superconducting systems with the FC state}\label{II}

Let us begin by considering the superconducting state of
high-temperature superconductors within the framework of the FC
theory \cite{ks,phys_rep}. Our consideration allows us to
demonstrate that the quasiparticles are well-defined excitations
and in the superconducting state of high-$T_c$ superconductors the
elementary excitations are Bogoliubov quasiparticle (BQ), and the
FC theory of the superconducting state is in general sense similar
to the Bardeen-Cooper-Schrieffer theory \cite{cryst24}. It was
observed that in high-$T_c$ the quasiparticles are well-defined
excitations and in the superconducting state the elementary
excitations are Bogoliubov quasiparticle (BQ), i.e. the excitations
are similar to quasiparticle excitations of conventional
superconductors \cite{bcs,til,genn,mat,npbq,prlq}. Thus,
unconventional superconductors exhibit the same scaling behavior of
the condensation energy $E_{\Delta}/\gamma$ as conventional
superconductors, while the Homes' law, see Eq. \eqref{home}, is
valid in both cases \cite{prb2015}. As we shall see, the
applicability of the BCS formalism  in describing the
superconducting state is closely related to the deformation of flat
band by the superconducting phase transition
\cite{shag1,ms,khod97,jetplbq,epl22}. Note that a number of
properties, such as the maximum value of the superconducting gap
$\Delta_1$, see Eq. \eqref{TC}, high density of states and other
exotic properties, go beyond the scope of the BCS theory
\cite{phys_rep,jetplbq,book_20}.

At $T<T_c$ the thermodynamic potential $\Omega$ of high $T_c$ is
determined by Equation, see e.g. \cite{lanl,til,genn})
\begin{equation}\label{1*}
\Omega=E_{gs}-\mu N-TS,
\end{equation}
In Eq. \eqref{1*} $N$ is the number density of quasiparticles that
coincides with the number density of particles, $\mu$ is the
chemical potential, $S$ is the entropy. The ground state energy
$E_{gs}[\kappa({\bf p}),n({\bf p})]$ of the electron liquid is an
exact functional of the order parameter of the superconducting
state $\kappa({ \bf p})$ and quasiparticle occupation numbers
$n({\bf p})$ \cite{phys_rep,pla98}. This energy is determined by
the well-known equation of the theory of weak coupling
superconductivity \cite{til,genn}
\begin{equation}\label{2*} E_{gs}\ =\
E[n({\bf p})]+\delta E_s.
\end{equation}
Here $E[n({\bf p})]$ is the exact Landau functional determining the
ground-state energy of normal Fermi liquid \cite{lanl,phys_rep},
and $\delta E_s$ is given by
\begin{equation}\label{2**}
\delta E_s=\int\lambda_0V({\bf p}_1,{\bf p}_2)\kappa({\bf p}_1)
\kappa^*({\bf p}_2)\frac{d{\bf p}_1d{\bf p}_2}{(2\pi)^4}.
\end{equation}
Here $\lambda_0V({\bf p}_1,{\bf p}_2)$ is the pairing interaction.
The quasiparticle occupation numbers
\begin{equation}\label{3*}
n({\bf p})=v^2({\bf p})(1-f({\bf p}))+u^2({\bf p})f({\bf p}),
\end{equation}
and at finite temperatures the order parameter $\kappa$ becomes
\begin{equation}\label{4*}
\kappa({\bf p})=v({\bf p})u({\bf p})(1-2f({\bf p})).
\end{equation}
While at $T=0$ the order parameter reads \cite{ksk}
\begin{equation}\label{17**}
\kappa({\bf p})=\sqrt{n_0({\bf p})(1-n_0({\bf p}))}.
\end{equation}
Here the coherence factors $u({\bf p})$ and $v({\bf p})$ obey the
normalization condition
\begin{equation}\label{5*}
v^2({\bf p})+u^2({\bf p})=1.
\end{equation}
The distribution function $f({\bf p})$ determines the entropy
\begin{equation}\label{6*}
S=-2\int\left[f({\bf p})\ln f({\bf p})+(1-f({\bf p}))\ln(1-f({\bf
p}))\right]\frac{d{\bf p}}{4\pi^2}.
\end{equation}
We will assume that the pair interaction $\lambda_0V({\bf p}_1,{\bf
p}_2)$ is weak and arises due to electron-phonon interaction.
Minimizing $\Omega$ with respect to $\kappa({\bf p})$ and using the
definition $\Delta({\bf p})=-\delta\Omega/\kappa({\bf p})$ , we
obtain the Equation , relating single-particle energy
$\varepsilon({\bf p})$ to the superconducting gap $\Delta({\bf p})$
\cite{til,genn}
\begin{equation}\label{7*}
\varepsilon({\bf p})-\mu\ =\Delta({\bf p})\frac{1-2v^2({\bf p})}
{2v({\bf p})u({\bf p})}.
\end{equation}
Single-particle energy $\varepsilon({\bf p})$ is determined by the
Landau equation
\begin{equation}\label{8*}
\varepsilon({\bf p})= \frac{\delta E[n({\bf p})]}{\delta n({\bf
p})}.
\end{equation}
Note that $E[n({\bf p})]$, $\varepsilon[n({\bf p})]$, and the
Landau amplitude
\begin{equation}\label{9*}
F({\bf p},{\bf p}_1)=\frac{\delta E^2[n({\bf p})]}{\delta n({\bf
p})\delta({\bf p}_1)}
\end{equation}
are exact equations \cite{pla98}. Minimizing $\Omega$ with respect
to $f({\bf p})$ and after some algebra, we obtain the well-know
equation for the superconducting gap $\Delta({\bf p})$
\begin{equation}\label{10*}
\Delta({\bf p})=-\frac{1}{2}\int\lambda_0 V({\bf p},{\bf p}_1)
\frac{\Delta({\bf p}_1)}{E({\bf p}_1)}(1-2f({\bf p}_1))\frac{d{\bf
p}_1}{4\pi^2}.
\end{equation}
Here the excitation energy $E({\bf p})$ is defined by the
Bogoliubov quasiparticles
\begin{equation}\label{11*}
E({\bf p})=\frac{\delta (E_{gs}-\mu N)}{\delta f({\bf p})}=
\sqrt{(\varepsilon({\bf p})-\mu)^2+\Delta^2({\bf p})}.
\end{equation}
Coherence factors $v({\bf p})$, $u({\bf p})$ and the distribution
function $f({\bf p})$ is given by the usual equations
\begin{equation}\label{12*}
v^2({\bf p})=\frac{1}{2}\left(1-\frac{\varepsilon({\bf
p})-\mu}{E({\bf p})}\right); u^2({\bf
p})=\frac{1}{2}\left(1+\frac{\varepsilon({\bf p})-\mu}{E({\bf
p})}\right),\end{equation}
\begin{equation}\label{13*}
f({\bf p})=\frac{1}{1+\exp(E({\bf p})/T)}.
\end{equation}
Equations \eqref{7*}---\eqref{13*} are the traditional equations of
the BCS theory \cite{bcs,til,genn}, defining the superconducting
state with BQ and the maximum value of the superconducting gap
$\Delta_1\ \sim 10^{-3} \varepsilon_F$ provided that it is assumed
that the compound in question has not been subjected to FCQPT.

At $T=0$ Eq. \eqref{10*} becomes \cite{til,genn}
\begin{equation}\label{100*}
\Delta({\bf p})=-\frac{1}{2}\int\lambda_0 V({\bf p},{\bf p}_1)
\frac{\Delta({\bf p}_1)}{E({\bf p}_1)}\frac{d{\bf p}_1}{4\pi^2}.
\end{equation}
Here we take into account that
\begin{equation}\label{101*}
2\kappa({\bf p})=\frac{\Delta({\bf p})}{E({\bf p})}.
\end{equation}
Taking into account Eqs. \eqref{100*}, \eqref{101*} and
\eqref{17**}, we return to Eq. \eqref{TC}. Thus, we see that the
BSC formalism works, but the existence of flat bands gives rise to
new features. On the other hand, as we will see in Section
\ref{III}, superconductivity deforms the flat band and makes the
effective mass finite, restoring the BCS formalism again
\cite{phys_rep}. These facts reveal new similarities between
unconventional and conventional superconductors, as shown in
Section \ref{unpr}.

\section{Common scaling of conventional and high-$T_c$
superconductors}\label{III}

\begin{figure}[!ht]
\begin{center}
\includegraphics[width=0.5\textwidth]{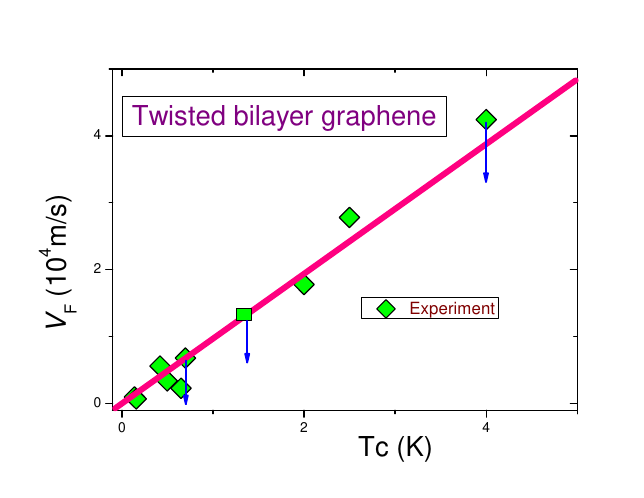}
\end{center}
\vspace*{-0.8cm} \caption{Experimental results (shown by the
squares) for the average Fermi velocity $V_F$ versus the critical
temperature $T_c$ for graphene (MATBG) \cite{mac}. The downward
arrows depict that $V_F\leq V_0$, with $V_0$ is the maximal value
shown by the red square. Theory is displayed by the solid straight
line.} \label{fig05}
\end{figure}

As can be seen from equations \eqref{7**} and \eqref{7*}, the
system is divided into two subsystems of quasiparticles: the first
subsystem in the range $(p_f-p_i)$ is characterized by
quasiparticles with an effective mass $M^* _{ FC}\propto
1/\Delta_1$, and the second is occupied by quasiparticles with a
finite mass $M^*_L$ and momenta $p<p_i$ \cite{phys_rep,epl22}. The
energy scale $E_0$ which defines the region occupied by
quasiparticles with the effective mass $M^*_{FC}$ Quasiparticles of
the effective mass $M^*_{FC}$ determine the energy scale $E_0$
\begin{equation}\label{17*} E_0=
\varepsilon({\bf p}_f)-\varepsilon({\bf p}_i) \simeq 2
\frac{(p_f-p_F)p_F}{M^*_{FC}}\ \simeq\ 2\Delta_1.
\end{equation}
As can be seen from Eq. \eqref{10*}, the superconducting gap
depends on the single-particle spectrum $\varepsilon({\bf p})$.
From Eq. \eqref{7*} it follows that $\varepsilon({\bf p})$ depends
on $\Delta({\bf p})$, since at $\Delta_1\to0$ the spectrum becomes
flat, and $M^*_{FC}$ becomes \cite{epl22}
\begin{equation}\label{15*}
M^*_{FC}\sim p_F\frac{p_f-p_i}{2\Delta_1}.
\end{equation}
As seen from Eq. \eqref{15*}, the effective mass and the density of
states $N(0)\propto M^*_{FC}\propto 1/\Delta_1$ are finite and
constant at $T<T_c$ \cite{ms,epl22}. At $T\to0$ and $\lambda_0\to
0$ the density of states near the Fermi level tends to infinity.
Thus, we arrive at the result that contradicts the BCS theory, and
follows from Eq. \eqref{15*}
\begin{equation}\label{15}
N(0)\propto M^*_{FC}\propto 1/\Delta_1\propto 1/T_c\propto 1/V_F,
\end{equation}
where $V_F\propto p_F/M^*_{FC}$ is the Fermi velocity
\cite{ms,mac,epl22}, see Fig. \ref{fig05}.

\begin{figure}[!ht]
\begin{center}
\vspace*{-0.8cm}
\includegraphics[width=0.5\textwidth]{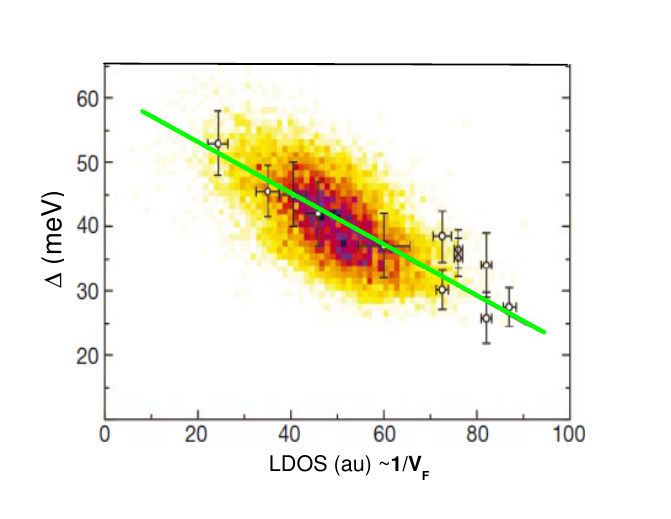}
\end{center}
\vspace*{-0.8cm} \caption{The figure is adapted from \cite{nat01},
and shows experimental dependence of the superconducting gap
$\Delta$ versus the integrated local density of states collected on
the high-$T_c$ superconductor $\rm Bi_2Sr_2CaCu_2O_{8+x}$. Here
$\rm x$ is oxygen doping concentration. Darker color represents
more data points with the same integrated local density of states
and the same size gap $\Delta$ \cite{nat01}. The straight blue line
shows average value $\Delta$ versus the integrated local density of
states.} \label{Fig2aa}
\end{figure}

The measurements of $V_F$ as a function of $T_c$ \cite{mac} are
shown in Fig. \ref{fig05}, and are in good agreement with  Eq.
\eqref{15}. Thus, our theoretical prediction
\cite{ms,phys_rep,shag} is in good agreement with observations
\cite{mac}. As can be seen from Fig. \ref{Fig2aa}, this unusual
behavior $T_c\propto\Delta_1\propto 1/N(0)$ is observed in
measurements on the unconventional superconductor $\rm
Bi_2Sr_2CaCu2O_{8+x}$, where $x$ is the oxygen doping concentration
and LDOS is the local integrated density of states
\cite{epl22,nat01}. Note that $V_F\to 0$ and also $T_c\to 0$, as
can be seen from Fig. \ref{fig05} and \ref{Fig2aa}. This result
shows that the flat band is broken by a finite value of $\Delta_1$
and has a finite slope that makes $V_F\propto T_c$, as can be seen
from Figs. \ref{fig05} and \ref{Fig2aa}. Indeed, it is clear from
these Figures that the maximum critical temperatures $T_c$ do not
correspond to the minima of the Fermi velocity $V_F$, as would be
the case in any BCS-type theory \cite{mac,nat01}. As we will see
below, another unusual behavior, namely the overall universal
scaling of $E_{\Delta}/\gamma$ of both conventional and high-$T_c$
superconductors \cite{prb2015}, is also related to Eq. \eqref{15}
and is explained within the framework of the FC theory.

We conclude that, in contrast to the traditional BCS theory of
superconductivity, the single-particle spectrum of
$\varepsilon({\bf p})$ strongly depends on the superconducting gap
and we need to solve Eqs. \eqref{8*} and \eqref{10*} in a
consistent manner. Let us suppose that Eqs. \eqref{8*} and
\eqref{10*} are solved and the effective mass $M^*_{FC}$ is fixed.
Now we can fix $\varepsilon({\bf p})$ by choosing the effective
mass $M^*$ of the system under consideration to be $M^*_{FC}$ and
then solving Eq. \eqref{10*} in the same way as is done in the case
of the conventional theory of superconductivity \cite{bcs}. As a
result, it is seen that the superconducting state is characterized
by BQ with dispersion determined by Eq. \eqref{11*}. Coherence
coefficients $v$, $u$ are determined by Eq. \eqref{12*}, and the
normalization condition \eqref{5*} is satisfied. We conclude that
the observed features are consistent with the BQ behavior according
to experimental facts \cite{mat,nakam}. This observation suggests
that a superconducting state with FC is similar to BCS and implies
the basic reliability of BCS formalism when describing a
superconducting state in terms of BQ. This is exactly the case that
was observed experimentally in high-$T_c $ cuprates, see e.g.
\cite{mat,npbq}.

\begin{figure}
\begin{center}
\includegraphics [width=0.47\textwidth]{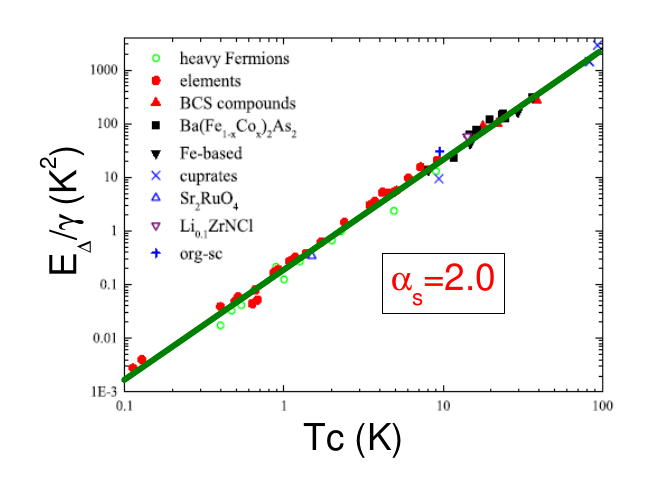}
\end{center}
\caption{Condensation energy $E_\Delta/\gamma\propto T_c^{2}$
divided by the specific heat $\gamma$ as a function of $T_c$ for a
wide range of superconductors, with the slope $\alpha_{s}=2$
\cite{prb2015}, see Eq. \eqref{30}. Deviations from the line of
best fit, spanning six orders of magnitude for $E_\Delta/\gamma$
and almost three orders of magnitude for $T_c$, are relatively
small.}\label{fig1}
\end{figure}

As a result, we obtain the usual BCS result for the superconducting
condensation energy $E_{\Delta}$, which is valid for both
conventional superconductors and unconventional superconductors
\cite{cryst24}
\begin{equation}\label{30}
\Delta E_{FC}/\gamma\sim E_{\Delta}/\gamma\sim
\frac{N(T)\Delta_1^2}{\gamma(T)}\sim \Delta_1^2\sim T_c^2.
\end{equation}
Here $N(T)$ and $\gamma(T)$ are the density of states and the
Sommerfeld coefficient, respectively. The factors $N(T)$ and
$\gamma(T)$ strongly depend on temperature $T$ in the FC theory,
and $\Delta_1$ is the maximum value of the superconducting gap.
However, $M^*(T)\propto N(T)\propto \gamma(T)$ \cite{phys_rep}, and
we have $E_{\Delta}/\gamma\sim T_c^2$. From Fig. \ref{fig1} it is
evident that Eq. \eqref{30} corresponds to the experimental facts
\cite{prb2015}. Taking into account that the BQ of non-traditional
high-temperature superconductors within the FC theory coincide with
the BQ of conventional superconductors and Eq. \eqref{15*}, we come
to the conclusion that the condensation energy $E_{\Delta}/\gamma$,
given by Eq. \eqref{30}, has a universal form, valid both in the
case of conventional superconductors and high-temperature
superconductors. To test this conclusion, we compare our
theoretical result with experimental facts \cite{prb2015}. Figure
\ref{fig1} shows the scaling of the condensation energy
$E_{\Delta}$ as a function of $T_c^2$ on a double logarithmic
scale. From Fig. \ref{fig1} it is evident that the universal
scaling $E_{\Delta}/\gamma\propto T_c^2$ is valid for all
superconductors, both conventional and non-conventional with high
$T_c$. This universal scaling behavior holds over nearly seven
orders of magnitude for $E_{\Delta}/\gamma$ and three orders of
magnitude for $T_c$ \cite{prb2015}. This observation is not
surprising since, as we saw above, high-$T_c$ superconductors have
the same BQ as ordinary ones, and the shape of the corresponding
band correlates with their $T_c$, as follows from Eq. \eqref{15}.
Note that due to the strong influence of the pseudogap state on the
properties of non-traditional superconductors, such as the density
of states, heat capacity, and even the true value of $T_c$ is
unclear, only optimally doped samples were considered
\cite{prb2015,loram}. Thus, the FC theory allows us to justify Eq.
\eqref{30}, which describes superconductivity that goes far beyond
the weak coupling regime and is applicable to both conventional and
unconventional strongly correlated superconductors.

\subsection{Heavy fermion metals as superconductors}

In passing, we note that the deviation of the effective mass $M^*$,
obtained from experimental data on heat capacity, from the bare
mass $M$ can be significant, as, for example, in HF metals. As we
will see, this deviation plays an important role in the value of
$T_c$.  In this connection, it is worth noting that from Eq.
\eqref{15} it directly follows that HF metals, having extremely
high values of the effective mass, have extremely low $T_c$
\cite{pag1,FlB,SHVM}. Indeed, the maximum value of $T_c$ is
achieved in the HF metal $\rm CeCoIn_5$ at $T_c\simeq 2.3$ K. This
observation agrees well with the experimental facts presented in
Fig. \ref{fig05} and \ref{Fig2aa}. Indeed, from these Figures it is
evident that $T_c\propto V_F\propto 1/M^*$.

\section{Linear temperature dependence of resistivity and universal
scattering rate}\label{IV}

\begin{figure} [! ht]
\begin{center}
\includegraphics [width=0.55\textwidth] {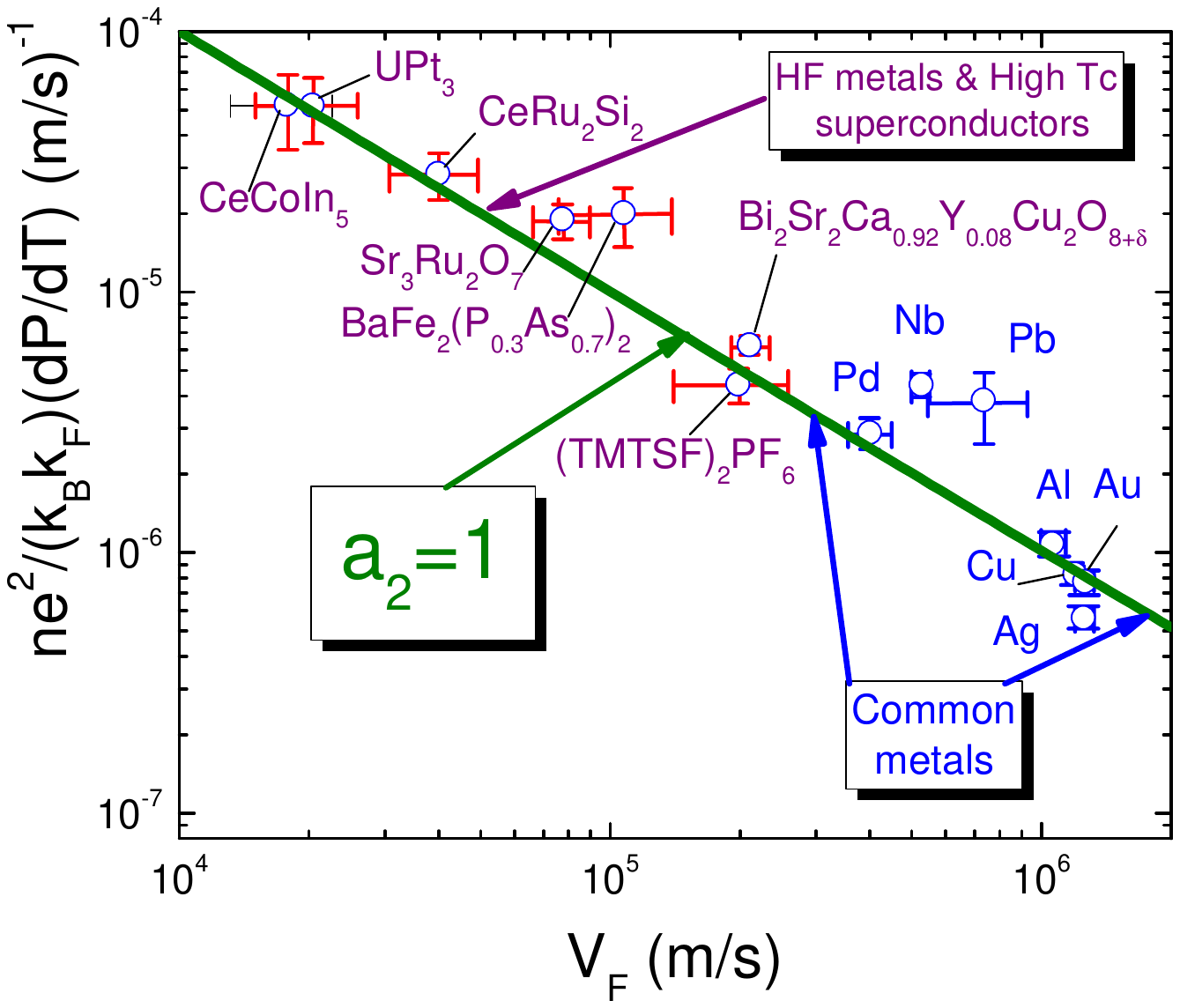}
\end {center}
\caption{Scattering rates of different strongly correlated metals
like HF metals, high-$T_c$ superconductors, organic metals, and
conventional metals \cite{bruin}. All these metals exhibit
$\rho(T)\propto T$ and demonstrate two orders of magnitude
variations in their Fermi velocities $v_F$. The parameter
$a_2\simeq 1$ corresponds to the Planckian limit, and gives the
best fit shown by the solid line, see Eq. \eqref{vf}. The region
occupied by the common metals is displayed by the two arrows, and
the arrow shows the region of strongly correlated metals}
\label{Sc1}
\end{figure}

When studying the linear-temperature resistivity of completely
different metals, such as HF metals, high-$T_c$ superconductors and
ordinary metals, at different temperatures $T$, the universality of
their fundamental physical properties was revealed
\cite{phys_rep,quasi,book,khod2012}. On the one hand, at low $T$
the linear $T$-resistivity
\begin{equation}
\rho(T)=\rho_0+AT,\label{res}
\end{equation}
observed in many strongly correlated compounds such as high-$T_c$
superconductors and HF metals is close to their QCP and hence
exhibits quantum criticality and a new state of matter. Here
$\rho_0$ is the residual resistivity and $A$ is a $T$-independent
coefficient. Explanations based on quantum criticality for the
$T$-linear resistivity have been given in literature, see e.g.
\cite{varma,varma1,phill,phill1,DP,khod2012} and references
therein. On the other hand, at room temperatures the $T$-linear
resistivity is exhibited by conventional metals such as $\rm Al$,
$\rm Ag$ or $\rm Cu$. In case of a simple metal with a single Fermi
surface pocket the resistivity reads $e^2n\rho=p_F/(\tau v_F)$,
\cite{trio} where $e$ is the electronic charge, $\tau$ is the
lifetime, $n$ is the carrier concentration, and $p_F$ and $v_F$ are
the Fermi momentum and  velocity respectively. We write the
lifetime $\tau$ (or the inverse scattering rate) of quasiparticles
in the form \cite{tomph,khod2012}
\begin{equation}\label{LT}
\frac{\hbar}{\tau}\simeq a_1+\frac{k_BT}{a_2},
\end{equation}
and obtain \cite{quasi}
\begin{equation}\label{vf}
a_2\frac{e^2n\hbar}{p_Fk_B}\frac{\partial\rho}{\partial
T}=\frac{1}{v_F},
\end{equation}
where $a_1$ and $a_2$ are $T$-independent parameters. The difficult
point for the theory is that experimental data confirm Eq.
\eqref{vf} for both strongly correlated metals (HF metals and
high-$T_c$ superconductors) and ordinary metals, provided that they
exhibit a linear dependence of their resistivity on temperature
\cite{bruin}, see Fig. \ref{Sc1}.

An examination of the literature data for most compounds with a
linear dependence $\rho(T)$ shows: The coefficient $a_2$ is always
close to unity, $0.7\leq a_2\leq 2.7$, despite the huge difference
in the absolute value of $\rho$, $T$ and the Fermi velocities
$v_F$, varying by two orders of magnitude \cite{bruin}. Thus, it
follows from Eq. \eqref{LT} that the $T$-linear scattering rate has
a universal form, $1/(\tau T)\sim k_B/\hbar$, regardless of the
various systems demonstrating a $T$-linear dependence with a
parameter included in Eq. \eqref{vf}, $a_2\simeq 1$,
\cite{bruin,quasi,book}. Indeed, this dependence manifests itself
(due to the electron-phonon mechanism) in ordinary metals at
temperatures above the Debye temperature, $T\geq T_D$. The same
dependence holds in strongly correlated metals, which are supposed
to be fundamentally different from ordinary ones. In the latter
substances, the linear dependence at their QCP and temperatures of
a few degrees Kelvin is supposed to come from excitations of
electronic origin, not from phonons \cite{bruin}. Note that in some
cuprates the scattering rate has a momentum and doping dependence
that is omitted from Eq. \eqref{vf} \cite{peets,french,alld}.
However, the fundamental picture described by Eq. \eqref{vf} is
convincingly confirmed by measurements of the resistivity of $\rm
Sr_3Ru_2O_7$ over a wide temperature range: at $T\geq 100$ K the
resistivity again becomes $T$-linear for all applied magnetic
fields, as it does at low temperatures and in the critical field
$B_{c}\simeq 7.9$ T, but with a coefficient $A$ lower than at low
temperatures \cite{bruin}. Thus, the same strongly correlated
compound exhibits the same resistance behavior in both the quantum
critical regime and the high-temperature regime, which allows us to
expect that similar physics governs the $T$-linear resistance,
despite the possible peculiarities of some compounds and ordinary
metals.
\begin{figure} [! ht]
\begin{center}
\includegraphics [width=0.55\textwidth]{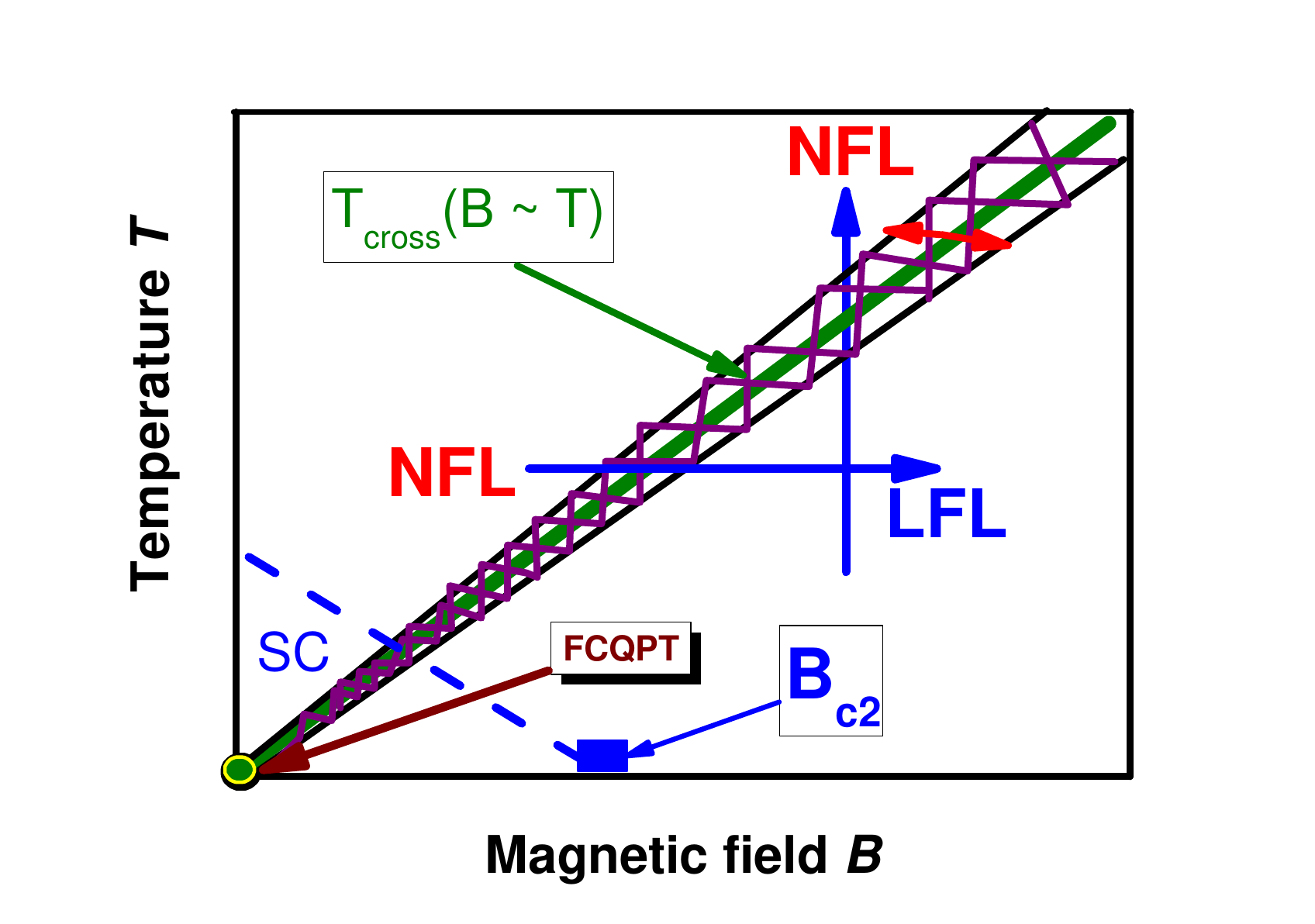}
\end{center}
\caption{Schematic $T-B$ phase diagram of a strongly correlated
Fermi system. The vertical and horizontal arrows crossing the
transition region (hatched area) depict the LFL-NFL and NFL-LFL
transitions at fixed $B$ and $T$, respectively. At $B<B_{c2}$ the
system is in its possible superconducting (SC) state, with $B_{c2}$
is shown by the solid line denoting a critical magnetic field
destroying the SC state. The hatched area with the solid curve
$T_{\rm cross}(B\sim T)$ represents the crossover region separating
NFL and LFL domains. A part of the crossover region is hidden in
the possible SC state.}\label{Fig2PD}
\end{figure}
As can be seen from Fig. \ref{Sc1}, this scaling relation spans two
orders of magnitude in $v_F$, which indicates the reliability of
the observed empirical law \cite{bruin}. This behavior is well
explained by the FC theory, since for both ordinary and strongly
correlated metals the scattering rate is determined by phonons
\cite{quasi}. In the case of ordinary metals at $T>T_D$ it is a
well-known fact that phonons make the main contribution to the
linear dependence of the resistivity, see e.g. \cite{trio}. Indeed,
semiclassical physics describes the $T$-linear dependence of the
resistivity of strongly correlated metals at $T>T_D$, since the
flat bands forming the QCP generate a transverse zero-sound mode
with the Debye temperature $T_D$ located in the region of quantum
criticality \cite{DP,khod2012,quasi}. Thus, the $T$-linear
dependence is formed by electron-phonon scattering in both ordinary
and strongly correlated metals. We conclude: It is the
electron-phonon scattering that leads to the almost complete
independence of the material of the lifetime $\tau$, which is
expressed as
\begin{equation}\label{planc}
 \frac{1}{\tau T}\sim \frac{k_B}{\hbar}.
\end{equation}
Note that in a magnetic field the system under consideration passes
from the NFL region to the LFL region, and both the flat bands and
the FC state are destroyed \cite{shag,phys_rep,book}, see the $T-B$
phase diagram in Fig. \ref{Fig2PD}. Therefore, the resistivity
$\rho(T)\propto T^2$ and magnetoresistance become negative, while
the residual resistivity $\rho_0$ drops sharply
\cite{khod2012,quasi,book}. Such behavior is in conformity with
experimental data, see e.g. the case of the HF metal $\rm CeCoIn_5$
\cite{pag1}.

\section{The superfluid density of overdoped copper
oxides}\label{cop}

\begin{figure}
\begin{center}
\includegraphics [width=0.47\textwidth]{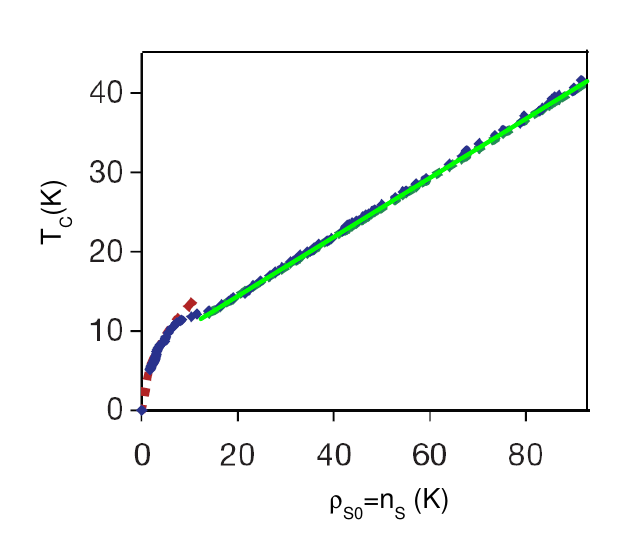}
\end{center}
\caption{The dependence of the critical temperature $Tc$ on the
superconducting density $n_s(T)=\rho_{s0}(T)$ at $T\to0$. The
experimental data are represented by the blue diamonds. The green
line is approximated $T_c=T_0+\alpha n_s$ and the red dashed line
is the fit to $T_c=\gamma\sqrt{n_s}$
\cite{pccp,bosovic}.}\label{fig1_aa}
\end{figure}

Our observation is that for the systems in question, the
approximate equality of the superconducting density $n_s$ of
electrons $n_s \simeq n_{el}$ what would be expected from
conventional BCS superconductors, should be replaced by the
inequality $n_s=n_{FC}\ll n_{el}$, where $n_{\rm FC}$ is particle
density in FC state \cite{pccp}. This implies that the main
contribution to $n_s$ comes from the FC state. Namely, the wave
function $\Xi$ describing the Cooper pairs as a whole, for the
system with a flat band behaves like $|\Xi|^2\propto n_s$ inside
the flat band and $|\Xi|^2 \simeq 0$ outside it. Near the FCQPT
point, where the FC fraction (flat band) is small in the momentum
region, $n_s$ can also become very small. Moreover, in this range,
$n_s$ does not depend on $n_{el}$, so one can expect that $n_s\ll
n_{el}$ \cite{pccp,khod97}.

It is worth noting that the first studies of the overdoped copper
oxides suggested that $n_s\equiv\rho_{s0}\ll n_{el}$, but this was
attributed to pair-breaking and disorder \cite{uemura,uem_n,bern1},
where $\rho_{s0}$ is the superfluid density at $T\to0$. It is seen
from Fig. \ref{fig1_aa} that recent measurements on ultra clean
samples of La$_{2-x}$Sr$_x$CuO$_4$ authenticate the result that
$n_s\ll n_{el}$ \cite{bosovic}. It is also relevant that the
observed high values of $T_c$ together with the linear dependence
$\rho_{s0}\propto T_c$ \cite{bosovic} of the resistivity are not
easily reconciled with the pair-breaking mechanism proposed for
dirty superconductors, see e.g. \cite{lee-h,khod2019}. One cannot
expect that such a mechanism would be consistent with high values
of $T_c$ and the increase of $T_c$ with doping. Overall, the data
support the FC theory, which suggests that the underlying physical
mechanism for both the unusual properties of overdoped copper
oxides and the asymmetry of tunnel conductivity is related to flat
bands \cite{phys_rep,pris2008,ks,vol,book} considered in Section
\ref{asc}.

\section{Asymmetrical conductivity}\label{asc}

Direct experimental studies of quantum phase transitions in
high-$T_c$ superconductors and HF metals are of great importance
for understanding the underlying physical mechanisms responsible
for their anomalous properties.  However, such studies are
difficult as the corresponding critical points are usually
concealed by the proximity to other phase transitions, most
commonly antiferromagnetic (AF) and superconducting (SC). Moreover,
extraordinary properties of resistivity in a magnetic field were
recently observed in graphene  having a flat band \cite{graph}
similar to high-$T_c$ superconductors and the HF metal $\rm
YbRh_2Si_2$ \cite{steg11,steg17}. Measuring and analyzing these
properties sheds light on the nature of the quantum phase
transitions in these substances. Very recently the scattering rate
has been measured in graphene. It turns out that it is located near
the universal value \cite{cao} given by Eq. \eqref{planc}, being in
accordance with the data shown in Fig. \ref{Sc1}. All these
experimental observations make graphene to be very interesting
material for revealing the underlying physics of strongly
correlated Fermi systems.

\begin{figure}[!ht]
\begin{center}
\includegraphics[width=0.5\textwidth]{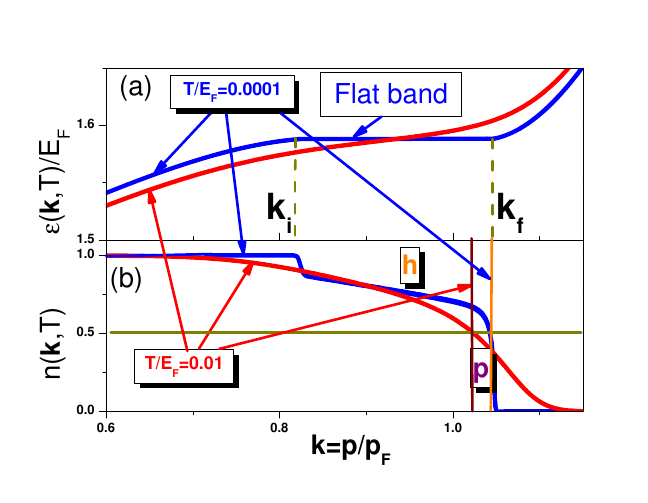}
\end{center}
\caption{Flat band induced by FC. The calculated single-particle
spectrum (a) and the quasiparticle occupation number (b) at small
but finite temperatures versus the dimensionless momentum
$k=p/p_F$, where $p_F$ is the Fermi momentum
\cite{pla_2007,phys_rep}. Temperature is measured in the units of
$E_F$. At $T=0.01E_F$ and $T=0.0001E_F$ the vertical lines show the
position of the Fermi level $E_F$ at which $n({\bf k},T)=0.5$ (see
the horizontal line in panel (b)). At $T=0.0001E_F$ (blue curve),
the single-particle spectrum $\varepsilon({\bf k},T)$ is almost
flat (marked "Flat band") in the range $k_f-k_i$ (with
$k_i=p_i/p_F$ and $k_f=p_f/p_F$ denoting respectively the
dimensionless initial and final momenta for FC realization). Thus,
the density of states $N_0\to\infty$ in the range $k_f-k_i$, and is
finite outside the range. The distribution function $n({\bf k},T)$
becomes more asymmetric with respect to the Fermi level $E_F$,
generating the NFL behavior and C invariance is broken. To
illuminate the asymmetry, the area occupied by holes in panel (b)
is labeled by h and that occupied by quasiparticles, by p.}
\label{Fig1C}
\end{figure}

\begin{figure} [! ht]
\begin{center}
\includegraphics [width=0.47\textwidth] {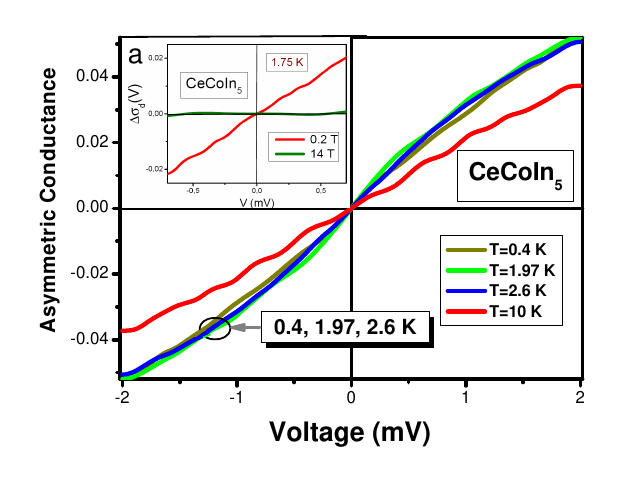}
\end {center}
\caption{The asymmetric part of tunneling conductivity
$\Delta\sigma_d(V)$ in $\rm CeCoIn_5$, extracted from the
experimental data \cite{steg2014}. At $T\leq 2.7$ K $\rm CeCoIn_5$
is in its pseudogap (PG) and superconducting states
\cite{steg2014}. At $T\leq 2.7$ K, as it is shown by both the ring
and the arrow, $\Delta\sigma_d(V)$ is temperature independent
\cite{phys_rep}. Inset a:  Asymmetric part $\Delta\sigma_d(V)$ of
the tunneling differential conductivity measured on $\rm CeCoIn_5$
and extracted from the experimental data \cite{park1}. Linear
dependence of $\Delta\sigma_d$ is shown by the straight line. The
asymmetric part disappears at $B=14$ T and $T=1.75$ K, with
$B_{c0}\simeq 5$ T.} \label{Fig2}
\end{figure}

Most of the experiments on HF metals and high-$T_c$ superconductors
explore their thermodynamic properties. However, it is equally
important to determine other properties of these systems, notably
quasiparticle occupation numbers $n(p,T)$ as a function of momentum
$p$ and temperature $T$. This is because the function $n(p,T)$ is
very helpful in the detection of time-reversal (T) and
particle-hole (C) symmetries violation, which are intimately
related to the NFL anomalies. We note that C-symmetry violation is
observed experimentally, see e.g. \cite{prx,step,chak} Scanning
tunneling microscopy \cite{harr,guy,zagos} and point contact
spectroscopy \cite{andr,tun,pla_2007,scr} are ideal tools for
exploring the effects of $C$ and $T$ symmetry violation, see e.g.
\cite{phmdpi}. When C and T symmetries are not conserved, the
differential tunneling conductivity (or resistivity) and dynamic
conductance are no longer symmetric functions of the applied
voltage $V$, since the particle-hole symmetry (C - symmetry) is
violated in systems with flat bands \cite{nphys19,phmdpi,scr}, see
Fig. \ref{Fig1C}. Indeed, if under the application of bias voltage
$V$, the current of negatively charged (charge $-e$) electrons
traveling from HF to an ordinary metal changes the sign to $+e$,
then current character and direction alters. Namely, now the
carriers become positively charged holes traveling in opposite
direction. At the same time, one arrives at the same current of
electrons if $V$ is changed to $-V$. The resulting asymmetric
differential conductivity
$\Delta\sigma_d(V)=\sigma_d(V)-\sigma_d(-V)$ becomes nonzero, as it
is seen from Fig. \ref{Fig2}. On the other hand, if time $t$ is
changed to $-t$ (but charge is kept intact), the current changes
its direction only. The same result can be achieved by $V\to -V$,
and we conclude that T symmetry is broken, provided that
$\Delta\sigma_d(V)\neq0$. Thus, if C or T symmetries are violated,
nonzero $\Delta\sigma_d(V)$ appears. Concurrently, the change of
both $e\to-e$ and $t\to-t$ returns the system to its initial state
so that CT symmetry is not broken. It is obvious, that the same
consideration is true for $\rho_d(V)=(\sigma_d(V))^{-1}$. Note that
the parity symmetry P is conserved and the well-known CPT symmetry
is not broken in the considered case. On the other hand, both T and
C symmetries remain intact in ordinary Fermi systems so that the
differential tunneling conductivity and dynamic conductance are
symmetric functions of $V$. This means that in normal metals at low
temperatures, no conductivity asymmetry is observed as in the case
of LFL \cite{phys_rep,scr,lanl}.

To obtain the tunneling conductivity (or point-contact spectra),
one first calculates the bias-dependent tunneling current $I(V)$
through the metallic point contact. This can be accomplished by
so-called Harrison method \cite{harr,guy,zagos}. This method
utilizes the fact that $I(V)$ is proportional to the Bardeen
transition probability \cite{bard}. On the other hand, if the
system hosts a FC, the asymmetry is strengthen by the asymmetry of
tunneling spectra as the density of states strongly depend on
$\varepsilon\simeq \mu$, see e.g. \cite{scr,book,nphys19}. The
situation becomes drastically different if a strongly correlated
Fermi system is placed near the FCQPT that engenders a flat band
\cite{ks,vol}, and violates both the C symmetry and T one
\cite{scr,phys_rep,book}. We note that as we have seen above, the
violation of C symmetry entails the violation of T symmetry.
Fig.~\ref{Fig1C} (a) illustrates the resulting low-temperature
spectrum $\varepsilon({\bf k},T)$. Fig.~\ref{Fig1C} (b), which
portrays the momentum dependence of the occupation numbers $n({\bf
k},T)$, shows that the flat band presence violates C symmetry. The
broken C symmetry results in the difference in areas of the regions
occupied by particles (labeled p) and holes (labeled h)
\cite{phys_rep,book,nphys19}. We note that a system in its
superconducting state exhibits the asymmetrical tunneling
conductivity near FCQPT. This is because in this case the C
symmetry remains broken in both the superconducting and the normal
states. It is seen from Fig. \ref{Fig2}, that this fact is in
accordance with experimental data
\cite{pla_2007,phys_rep,book,scr,nphys19}.

It is seen from Fig.~\ref{Fig1C} that at low temperatures the
electronic liquid of the system under consideration has two
components. One is an exotic component made up of heavy electrons
occupying a range of momenta $p_i < p < p_f$ surrounding the Fermi
volume near the former Fermi surface $p=p_F$. At $T=0$ this
component is characterized by the superconducting order parameter
$\kappa(p)=\sqrt{n(p)(1-n(p)}$ and generates the superconducting
density of electrons $n_s$. The other component consists of normal
electrons in the momentum range $0 \leq p \leq p_i$, which
maintains the LFL properties \cite{khod97,phys_rep}. This unusual
momentum distribution cannot be described within the standard BCS
theory. In particular, the density of paired charge carriers that
form the superfluid component is no longer equal to the total
particle density $n_{el}$ represented by paired and unpaired charge
carriers. This Leggett's theorem breaking is to be expected, since
both C - and T - invariance are violated in the NFL state of some
HF compounds \cite{book,phys_rep}.

In such cases it is again the occupation number $n({\bf p})$ that
is responsible for the asymmetric part of $\Delta \sigma_d(V)$,
since this function is not appreciably disturbed by the
superconductive pairing. This is because latter pairing is usually
weaker than the Landau interaction \cite{phys_rep}. As a result,
$\Delta \sigma_d(V)$ remains approximately the same below the
superconducting $T_c$, see Fig. \ref{Fig2}
\cite{phys_rep,tun,steg2014}. With raising temperatures the
asymmetry diminishes and finally goes to zero at sufficiently high
temperatures. Such behavior has been observed in measurements on
the HF metal $\rm CeCoIn_5$ \cite{park1}, and reported in
Fig.~\ref{Fig2}.
In a magnetic field $B$ at sufficiently low temperatures
$k_BT\lesssim \mu_BB$ ($\mu_B$ is Bohr magneton), the strongly
correlated Fermi system transits from the NFL to the LFL regime
\cite{phys_rep,pog2002}, see the inset (a), Fig. \ref{Fig2PD}.  As
we have seen above, the asymmetry of the  conductivity vanishes in
the LFL state \cite{tun,pla_2007,tun_k,phys_rep}. We surmise that
$\Delta \sigma_d(V)$ seen in Fig. \ref{Fig2} should vanish in the
normal state at sufficiently high magnetic fields applied along the
easy axis and low temperatures $k_B T << \mu_B (B-B_c)$ with the
critical field $B_c\simeq 5$ T. Under this condition the system
transits from the NFL to the LFL region in the phase diagram
\ref{Fig2PD}, acquiring the LFL properties with the resistance
$\rho$ becoming a quadratic function of temperature,
$\rho(T)\propto T^2$, \cite{phys_rep}. Note, that latter vanishing
has been predicted many years before the experimental observations
\cite{pla_2007,tun,tun_k}. It is worth noting that the
disappearance of the asymmetric part when a magnetic field is
applied is an important point, since the presence of the asymmetric
part can be demonstrated using a simple example, for example, a
diode, while the asymmetric part in this case does not disappear in
magnetic fields. Note that at $B=0$ the asymmetric part observed in
HF metals and high-$T_c$ superconductors can be explained by a
number of reasons, see e.g. \cite{feng}.

\section{General universal properties of heavy-fermion metals,
high-temperature superconductors and
conventional superconductors}\label{unpr}

\begin{figure}
\begin{center}
\includegraphics [width=0.55\textwidth]{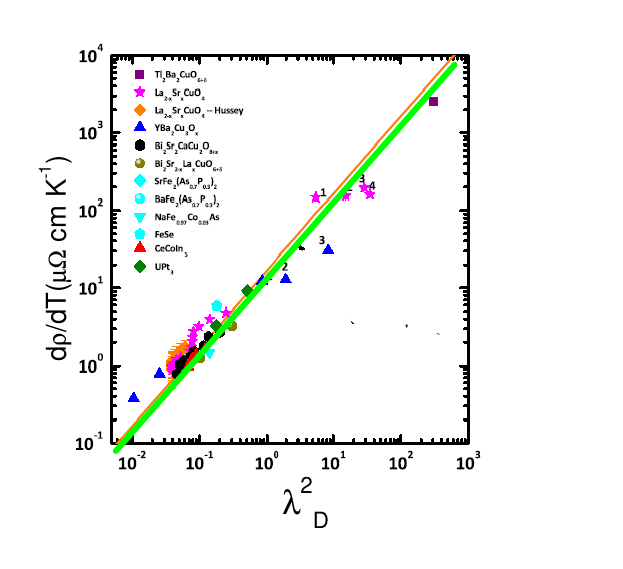}
\end{center}
\caption{ Double logarithm plot of $d\rho(T)/dT$ as a function of
$\lambda^2_D$ (see Eqs. \eqref{cs1} and \eqref{cs3}) for various
strongly correlated superconductors \cite{lin}. The green line
represents the theoretical calculations.}\label{fig5a}
\end{figure}

\begin{figure}
\begin{center}
\includegraphics [width=0.65\textwidth]{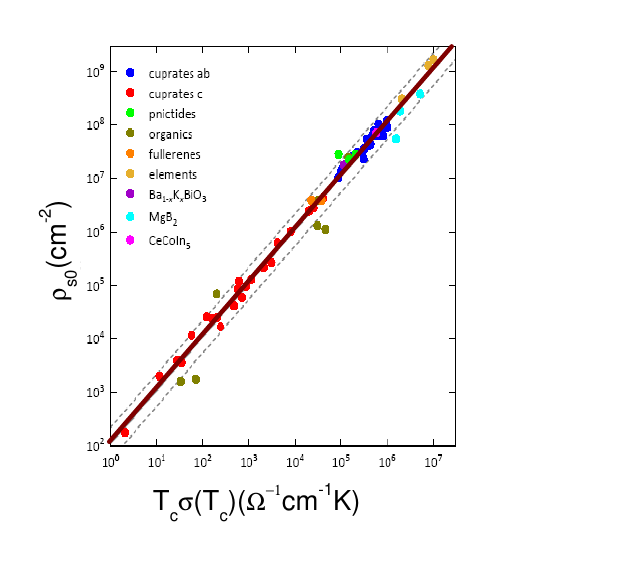}
\end{center}
\caption{Double logarithm plot of $\rho_{s0}$ as a function of
$T_c\sigma(T_c)$ for various superconductors \cite{donald}. The
wine line represents the theoretical calculations, see Eqs.
\eqref{cs4}.}\label{fig4a}
\end{figure}

It has been shown that the linear $T$-dependence of the resistivity
is an intrinsic property of cuprates and HF metals, associated with
a universal scattering rate (see Eq. \eqref{res})
\cite{cao,quasi,bruin,legr,khod2012,bruin,nakaj}. It is stated that
the behavior \eqref{res} is achieved when the scattering rate hits
the Planckian limit, given by Eq. \eqref{planc}, regardless the
origin of a scattering process \cite{cao,nakaj,legr}. On the other
hand, it is hardly possible that the linear resistance in $T$ of
ordinary metals will reach the Planck limit \cite{bruin}, see Fig.
\ref{Sc1}. Such a behavior is explained in Ref.
\cite{quasi,khod2012,book_20}. Moreover, HF metals and high-$T_c$
superconductors demonstrate the scaling behavior in a magnetic
field, pressure, etc. In magnetic fields, these compounds are
shifted from the NFL behavior to LFL one, as it is shown in Phase
diagram \ref{Fig2PD}, see e. g. \cite{phys_rep,nakaj}. All these
essential features are easily explained within the framework of the
FC theory \cite{ks,phys_rep,book}. As a result, we can safely
suggest that the main reason of the behavior given by Eq.
\eqref{planc} is phonon scattering, taking place in both strongly
correlated Fermi systems and ordinary metals and generating the
linear $T$-dependence of the resistivity \cite{khod2012,quasi} and,
as we shall see below, leads to Homes' law
\cite{homes,bruin,book_20}, see Eq. \eqref{res} and Figs.
\ref{Sc1}, \ref{fig5a} and \ref{fig4a}. Below, based on the results
of Section \ref{II}, we present a general derivation of Homes' law,
which becomes applicable to both non-traditional and conventional
superconductors.

Another experimental result \cite{lin} that provides insight into
the NFL behavior of strongly correlated Fermi systems is the
universal scaling relation, which can also be explained by FC
theory. The temperature dependence of $\rho(T)$ was measured for a
large number of superconducting compounds at $T>T_c$ \cite{lin}.
Among these were LSCO and the well-known HF compound CeCoIn$_5$;
see Table I of Ref.~\cite{lin}. Some rather unusual behavior was
observed: for all substances considered, $d\rho/dT$ shows a linear
dependence on $\lambda_D^2\propto M^*/n_s=M^*/\rho_{s0}$. All the
superconductors considered are of the London type, for which
$\lambda_D>> \xi_0$, where $\xi_0$ is the zero-temperature
coherence length (see, e.g., Ref.~\cite{pccp}).

The scaling relation
\begin{equation}\label{cs1}
\frac{d\rho}{dT} \propto \frac{k_B}{\hbar}\lambda_D^2
\end{equation}
has been shown to hold over several orders of magnitude of
$\lambda_D$, demonstrating its robustness \cite{lin}. At the phase
transition point $T=T_c$ the relation \eqref{cs1} leads to the
well-known Homes' law \cite{homes}
\begin{equation}\label{cs2}
\rho_{s0}\propto\frac{T_c}{\rho(T_c)}\propto\lambda_D^{-2}\propto
\frac{n_{el}}{M^*},
\end{equation}
where $\rho$ is the normal state resistivity. Within the framework
of the simple metal model, the  resistance $\rho$ can be expressed
through the microscopic parameters of the substance \cite{trio}:
\begin{equation} \label{cs2a}
e^2n\rho \simeq p_F/(\tau v_F)
\end{equation}
where $\tau$ is the quasiparticle lifetime, $n_{el}$ is the carrier
density. Taking into account that $p_F/v_F=M^*$, we arrive at Eq.
\cite{scr}
\begin{equation}\label{tau}
\rho=\frac{M^*}{n_{el}e^2\tau}.
\end{equation}
It is evident that Eq. \eqref{tau} is formally consistent with the
well-known Drude formula. Keeping these facts in mind and taking
into account the ratio \cite{khod2012,quasi}
\begin{equation}\label{cs33}
\frac{1}{\tau}=\frac{k_BT}{\hbar},
\end{equation}
we obtain
\begin{equation}\label{cs3}
\frac{d\rho}{dT}=\frac{M^*}{e^2n_{el}}\frac{k_B}{\hbar}\equiv
4\pi\lambda_D^2\frac{k_B}{\hbar},
\end{equation}
i.e. $d\rho/dT$ is indeed given by the expression \eqref{cs1}. Note
that the key point in our conclusion here is that the resistance
$\rho(T)\propto T$ \cite{khod2012,quasi}. As can be seen from Fig.
\ref{fig5a}, Eq. \eqref{cs3} describes the experimental facts well,
while the FC theory can explain the experimentally observed
universal scaling relation.

Now let us turn to Homes'scaling law
\cite{donald,kogan,valla,basov,boy,s2025}. It directly follows from
Eqs. \eqref{res} and \eqref{cs2} that the Homes' law is satisfied
\begin{equation}\label{cs4}
\rho_{s0}=C_0T_c\sigma(T_c),
\end{equation}
where $\sigma(T)=\frac{1}{\rho(T)}$. It is also shown that in the
BCS model, Homes scaling is applicable in both the dirty and clean
limits \cite{kogan,valla}
\begin{equation}\label{cs5}
\rho_{s0}=C_1\Delta\sigma(T_c),
\end{equation}
where $C_1\simeq\alpha (k_B/h)$ and $\alpha$ is a factor. On the
other hand, taking into account Eq. \eqref{30} and Fig. \ref{fig1},
we conclude that Eq. \eqref{home} is applicable to both
conventional and unconventional superconductors with $C_1\simeq
C_0$. As a result, we conclude that $\alpha\sim 1$, as in the case
of BCS \cite{kogan,s2025}. This possible deviation from unity of
$\alpha$ does not have a significant effect on the logarithmic
scale of Fig. \ref{fig4a} \cite{donald}. Equation \eqref{home} is
supported by experimental facts collected on various
superconductors \cite{donald,valla,basov}, as seen from Fig.
\ref{fig4a}. Thus, Eq. \eqref{home} shows the general behavior of
both conventional and unconventional superconductors.

A few remarks are in order here: It has been shown in
Ref.~\cite{pccp} that the effective mass $M^* \equiv M^*_{FC}$ and
the superfluid density equals to the carriers in the FC state only
$n_s\equiv n_{FC}<<n_{el}$. These observations are in a good
agreement with experimental results obtained on
quasi-two-dimensional overdoped single-crystal films of $\rm
La_{2-x}Sr_xCuO_4$ \cite{bosovic}, see the Section \ref{cop}. In
that case we obtain
\begin{equation}\label{cs3N}
\frac{d\rho}{dT}=\frac{M^*_{FC}}{e^2n_{s}}\frac{k_B}{\hbar}\equiv
4\pi\lambda_D^2\frac{k_B}{\hbar}.
\end{equation}
In that case at $T\to 0$ the superfluid density can be
$n_s<<n_{el}$, see the Section \ref{cop}, while $\rho(T)$ is given
be Eq. \eqref{tau}, we expect a deviation from Eq. \eqref{cs1}. In
the same way, (for $n_s<<n_{el}$) Homes' law, given by Eq.
\eqref{cs4}, is violated \cite{bosovic}.

It is important to emphasize that the approach presented here is
insensitive to the microscopic details (such as the specific
crystal structure and symmetry, its defect composition, etc.) of
the substances studied. This is explained by the fact that the FC
state is protected by its topological structure and therefore
represents a new class of Fermi liquids \cite{vol,book}. This
provides strong evidence that FC theory provides a robust
methodology for explaining universal scaling relationships such as
those found in experiments of Bo\^zovi\'c et al. \cite{bosovic} and
Hu et al. \cite{lin}. That is, the fermion condensation of charge
carriers in the substances under consideration is indeed the main
physical mechanism responsible for their exotic NFL properties.
This mechanism can be extended to a wide range of substances with
completely different microscopic characteristics, as discussed in
detail in Refs. \cite{phys_rep,book}.

\section{Conclusions}\label{conc}

The main idea of this review article is that if the electronic
spectrum of a substance has a dispersionless part, called flat
bands, then this effect is responsible for the measured properties,
which differ radically from the properties of known condensed
systems described by the Landau Fermi liquid theory. Flat bands
were predicted many years ago \cite{ks,ksk}, and now flat band
condensed matter physics is becoming one of the main areas of
modern physics. Arising from the topological phase transition, flat
bands give rise to the new state of matter that behaves
independently of the various microscopic details that characterize
these substances, such as crystal symmetry and defect structure.
The explanation for this finding is that FC occurs most readily in
materials containing flat bands. In our review, we have examined
the general properties of conventional and non-conventional
superconductors and demonstrated that, on the one hand, they
exhibit common features, but on the other hand, they exhibit
distinctly different behavior. We have emphasized that the overall
scaling behavior of superconductors is of crucial importance
because it reveals the essence of their physics. We have analyzed
recent challenging experimental observations listed below:

1. We have analyzed the common behavior of unconventional
high-$T_c$ and conventional superconductors and demonstrated that
the universal scaling of the condensation energy
$E_{\Delta}/\gamma=N(0)\Delta_1^2/\gamma$  applies equally to
conventional and unconventional high-$T_c$ superconductors. Our
explanation is based on the general property of superconductors:
Bogoliubov quasiparticles act in conventional and unconventional
superconductors, while the corresponding band is only deformed by
the unconventional superconducting state, making the effective mass
$M^*$ finite, see Eq. \eqref{15}. These observations suggest that
the unconventional superconducting state can be considered similar
to the BCS in many cases. Our theoretical observations agree well
with the experimental facts. On the other hand, the generally
accepted opinion assumes that flat bands are not deformed, i.e.
remain flat, when the corresponding superconducting state occurs,
see, for example, \cite{Aror,Tian}, which, as we have shown,
contradicts experimental facts.

2. We have considered in the framework of the FC theory that the
linear temperature dependence of the resistivity $\rho(T)\propto
T$, collected on such seemingly different substances as
high-temperature superconductors, HF compounds, and also ordinary
(i.e., obeying the LFL theory) metals, showing that the scattering
rate 1$/\tau$ of charge carriers reaches the Planck limit
\cite{bruin,legr,cao,nakaj}. We assume that the Planck limit arises
by chance, since normal metals exhibit the same scattering rate
behavior.

3. We also reviewed exciting experimental results on the
differential resistance (or conductivity) in a magnetic field of
the archetypal HF metal $\rm CeCoIn_5$, taken as an example. We
showed that in magnetic fields the systems under consideration
transition from NFL to LFL regimes, so that the asymmetry of tunnel
conductivity disappears. This is because the time-reversal (T) and
particle-hole (C) symmetries are restored in the LFL regime,
whereas they are broken in the NFL regime.

4. We have explained the observed facts that the density of the
superconducting electron $n_s$ of overdoped $\rm La_{2-x}Sr_xCuO_4$
does not coincide with the density of electrons $n_{el}$, so that
$n_{el}>>n_s$. We also explained how the high density of states is
consistent with the very low value of the superconducting
transition temperature $T_c$.

5. We have analyzed both $d\rho(T)/dT$ as a function of
$\lambda^2_D$ and of $\rho_{s0}$ as a function of $T_c\sigma(T_c)$
for various superconductors and shown that
$d\rho(T)/dT\propto\lambda^2_D$ and $\rho_{s0}\propto
T_c\sigma(T_c)$. We have analyzed the Homes' law and provide a
theoretical explanation of its general scaling applicable to
superconductors. We have also explained the case of violation of
the Homes' law. Overall, these scaling relationships lead to the
identification of fundamental laws of nature and reveal the essence
of superconductor physics.

Thus, our studies of these important experimental facts
convincingly indicate that the topological FCQPT is an integral
feature of many (at first glance different, i.e. possessing
different microscopic properties) strongly correlated Fermi systems
and can be considered as a universal reason for their exotic
non-Fermi-liquid properties, which cannot be observed in metals
described by the Landau Fermi liquid theory. Moreover, the FC
theory based on the topological FCQPT is able to explain the
complex behavior of strongly correlated Fermi systems represented
by unconventional superconductors and their relationship with
conventional superconductors, revealing the main features of
superconductor physics associated with the flat band phenomenon.

\section{Acknowledgement}
This work was supported by U.S. DOE, Division of Chemical Sciences,
Office of Basic Energy Sciences, Office of Energy Research, AFOSR.

\end{document}